\documentclass[12pt]{article}
\usepackage{amstex}
\usepackage{amssymb}
\usepackage{a4}
\usepackage{epsfig}
\begin{document}
\newcommand{\newc}{\newcommand}
\newc{\R}{$R$}
\newc{\charginom}{M_{\tilde \chi}^{+}}
\newc{\mue}{\mu_{\tilde{e}_{iL}}}
\newc{\mud}{\mu_{\tilde{d}_{jL}}}
\newc{\barr}{\begin{eqnarray}}
\newc{\earr}{\end{eqnarray}}
\newc{\beq}{\begin{equation}}
\newc{\eeq}{\end{equation}}
\newc{\ra}{\rightarrow}
\newc{\lam}{\lambda}
\newc{\eps}{\epsilon}
\newc{\gev}{\,GeV}
\newc{\tev}{\,TeV}
\newc{\eq}[1]{(\ref{eq:#1})}
\newc{\eqs}[2]{(\ref{eq:#1},\ref{eq:#2})}
\newc{\etal}{{\it et al.}\ }
\newc{\eg}{{\it e.g.}\ }
\newc{\ie}{{\it i.e.}\ }
\newc{\nonum}{\nonumber}
\newc{\lab}[1]{\label{eq:#1}}
\newc{\dpr}[2]{({#1}\cdot{#2})}
\newc{\gsim}{\stackrel{>}{\sim}}
\newc{\lsim}{\stackrel{<}{\sim}}

\begin{titlepage}
\begin{flushright}
{ETHZ-IPP  PR-97-01} \\
{May 5, 1997}\\
\end{flushright}
\begin{center}
{\bf \LARGE
Towards a Precise Parton Luminosity \\
Determination at the CERN LHC}
\end{center}

\smallskip \smallskip \bigskip
\begin{center}
{\Large M. Dittmar, F. Pauss and D. Z\"{u}rcher}
\end{center}
\bigskip

\begin{center}
Institute for Particle Physics (IPP), ETH Z\"{u}rich, \\
CH-8093 Z\"{u}rich, Switzerland
\end{center}

%\begin{center}
%{\LARGE Draft 3.0}
%\end{center}

\begin{abstract}
\noindent A new approach to determine the LHC luminosity 
is investigated. 
Instead of employing the proton--proton luminosity
measurement, we 
suggest to measure directly the parton--parton luminosity.
It is shown that the electron and muon 
pseudorapidity distributions, originating from the decay of 
$W^{+}$, $W^{-}$ and $Z^{0}$ bosons produced  
at 14 TeV pp collisions (LHC),
constrain the $x$ distributions
of sea and valence quarks and antiquarks
in the range from $\approx 3 \times 10^{-4}$ to 
$\approx 10^{-1}$ at a $Q^{2}$ of about $10^{4}$ GeV$^{2}$.
Furthermore, it is demonstrated that, once the quark and antiquark
structure functions are constrained from the 
$W^{\pm}$ and $Z^{0}$ production dynamics, 
other $q\bar{q}$ related scattering processes at the 
LHC like $ q\bar{q} \rightarrow W^{+}W^{-}$ can be predicted 
accurately. Thus, the lepton pseudorapidity distributions 
provide the 
key to a precise parton luminosity monitor at the LHC,
with accuracies of $\approx \pm 1$\%
compared to the so far considered  
goal of $\pm$5\%.  
\vspace{2cm}
\end{abstract}

\begin{center}
{\it submitted to Physical Review D }
\end{center}
\end{titlepage}

\section{Introduction}
Interpretations of essentially all 
proposed measurements at the LHC, 
CERN's 14 TeV proton--proton collider project,
require a good knowledge of the 
parton distribution functions at the relevant $Q^{2}$
and the collected integrated luminosity.
Both omnipurpose experiments, ATLAS \cite{ATLAS}
and CMS \cite{CMS}, consider a luminosity
accuracy of $\pm$5\% as their goal \cite{lumerror}.

The traditional methods to determine the 
proton--proton luminosity are size and intensity
measurements of the beams at the interaction point, 
as well as event rates of processes with previously measured 
or calculable cross sections like elastic 
proton--proton scattering \cite{ppelas} 
and QED processes like $pp \rightarrow pp e^{+}e^{-}$ \cite{ppee}.
Unfortunately, clean measurements of the 
above processes, especially at the high luminosity phase of LHC,  
are very difficult. Consequently it is not obvious 
that the proton--proton luminosity can be measured with  
a $\pm$5\% accuracy. 
 
Assuming that the proton--proton luminosity is measured, different
experimentally observed cross sections 
are compared with theoretical calculations, using  
parton distribution functions, 
f($x, Q^{2}$), where $x$ is the 
fractional parton momentum ($x = p_{\rm parton}/E_{\rm beam}$) of 
the relevant types of valence and sea quarks (or antiquarks)
and gluons at the considered $Q^{2}$ of the reaction.
These parton distributions are determined from 
experimental observables in lepton--hadron scattering 
(DIS processes from fixed target and HERA experiments) and  
Drell--Yan lepton pair production processes 
at hadron colliders \cite{hmsr}.
These results, obtained at different 
$Q^{2}$, have then to be extrapolated to the relevant $Q^{2}$ scale
of the studied process. While the $x$ distributions of 
the valence quarks are now quite well constrained,
uncertainties for the $x$ distributions of sea quarks and antiquarks 
and gluons remain important.
As a result of these structure function 
uncertainties, total cross section predictions
of $W^{+}$, $W^{-}$ and $Z^{0}$ boson production  
at 14 TeV pp collisions (LHC), vary currently between 10--20\%.
Even though the experimental errors are expected to 
decrease considerable during the next years,  
cross section uncertainties related to structure functions 
will remain important.
These uncertainties, combined 
with the unknown contributions 
from higher order QCD corrections, are 
usually considered to limit  
the use of the reaction $pp \rightarrow W^{\pm} (Z^{0})$ 
as an absolute proton--proton luminosity monitor 
at very high center of mass energies \cite{sdc}.

The above problems result in luminosity uncertainties,
which are larger than the considered goal of 
a $\pm$5\% proton--proton luminosity accuracy \cite{lumerror}. 
Consequently, current estimates for 
the achievable accuracies of some measurements at the LHC   
appear to be somewhat depressing. This is especially the case 
when these uncertainties are compared with 
the possible small statistical errors for many 
LHC measurements, the current knowledge 
of quark and lepton couplings to the 
$W^{\pm}$ and $Z^{0}$ boson or the already  
achieved accuracies in high energy $e^{+}e^{-}$ experiments. 

As a solution for the above problem,  
we propose a new approach to measure the LHC
luminosity. This approach is based on:
%\begin{itemize}
\begin{enumerate}
\item
Experiments at the LHC will study the interactions 
between fundamental constituents of the proton, the quarks and 
gluons at energies where these partons can be considered as 
quasi free. Thus, the important quantity is the 
parton--parton luminosity 
at different values of $x_{\rm parton}$ \cite{qcdcoll}
and not the   
traditionally considered proton--proton luminosity.
\item
Assuming collisions of essentially free partons,
the production of weak bosons, $u \bar{d} \rightarrow W^{+}
\rightarrow \ell^{+} \nu$,
$d \bar{u} \rightarrow W^{-}
\rightarrow \ell^{-} \bar{\nu}$
and $u \bar{u} (d \bar{d}) \rightarrow Z^{0} 
\rightarrow \ell^{+}\ell^{-}$ are in lowest order understood
to at least a percent level.
Cross section uncertainties from higher order QCD 
corrections are certainly larger, but are obviously 
included in the measured weak boson event rates. 
Similar higher order QCD corrections to other 
$q\bar{q}$ scattering processes 
at different $Q^{2}$, like $q\bar{q} \rightarrow W^{+}W^{-}$,
can be expected. Thus, 
assuming that the $Q^{2}$ dependence can in principle
be calculated, very accurate theoretical predictions
for cross section ratios like 
$\sigma(pp \rightarrow W^{+}W^{-})$/$\sigma(pp \rightarrow W^{\pm})$
should be possible. 
\item
It is a well known fact that 
the $W^{\pm}$ and $Z^{0}$ production rates at the LHC, 
including their leptonic branching ratios into electrons 
and muons, are huge and provide relatively clean  
and well measurable events with isolated leptons.
With the well known $W^{\pm}$ and $Z^{0}$ masses,
possible $x$ values of quarks and antiquarks
are constrained from
$M_{W^{\pm}, Z^{0}}^{2} = s x_{q} x_{\bar{q}}$
with $s = 4 E_{beam}^{2}$. The product
$x_{q} x_{\bar{q}}$ at the LHC ($\sqrt{s} = 14$ TeV) is thus 
fixed to $\approx 3 \times 10^{-5}$. 
Thus, the rapidity distributions of the weak bosons 
are directly related to the fractional momenta $x$ 
of the quarks and antiquarks.
Consequently, the observable pseudorapidity distributions of the 
charged leptons from the decays of $W^{\pm}$ and $Z^{0}$ bosons
are also related to the $x$ distributions of 
quarks and antiquarks.
The shape and rate of the lepton pseudorapidity distributions
provide therefore the key to precisely constrain the 
quark and antiquark structure functions 
and their corresponding luminosities.   
\end{enumerate}
%\end{itemize}
The aim of this paper is to demonstrate the feasibility 
of this approach and thus improve the luminosity measurement 
at the LHC for quark--antiquark related scattering processes. 
It will be shown that the dynamics of the single 
weak boson production at the LHC allow to constrain  
the $q,\bar{q}$ structure functions, the corresponding 
parton luminosities and therefore also the cross sections of
other $q\bar{q}$ related processes. 
Finally, we suggest that a similar approach to gluon related 
scattering processes might eventually also lead to similar 
accuracies for the $x$ distribution of gluons.
 
\section {Event rates and the selection of $pp \rightarrow$  
$W^{+}$, $W^{-}$ and $Z^{0}$}

The production of $pp \rightarrow$  
$W^{+}$, $W^{-}$ and $Z^{0}$ and their identification 
using the leptonic 
decays have been discussed extensively 
in the literature \cite{lhcwz}. 
In particular, these reactions 
provide clean sources of isolated high $p_{t}$ electrons
or muons, and due to their high rate, 
are often considered as a clean and
excellent calibration tool at the LHC \cite{calib}.  
However, previous studies concluded that their  
use as a luminosity monitor is limited to 
relative luminosity 
measurements only \cite{sdc}. The reason for these pessimistic 
conclusions is based on
the predicted cross section variations using different 
sets of structure functions \cite{mrsa}. The size of these 
cross section variations is as large as 10--20\% as 
can be seen from table 1. The cross section predictions 
as well as the following simulation results 
are obtained using the PYTHIA Monte Carlo program \cite{pythia}. 

These cross section variations
for single $W^{\pm}$, $Z^{0}$ production
are strongly correlated with the cross section  
predictions for other $q\bar{q}$ related processes. 
As an example, the corresponding cross sections for the reaction 
$q\bar{q} \rightarrow W^{+}W^{-}$ are also given in table 1.
Thus, even without looking at further details, 
the uncertainties for multi boson production cross sections
at the LHC are reduced to about 5-10\% if event rates 
are estimated relative to  
the production rates of single $W^{\pm}$, $Z^{0}$ events. 
Furthermore such relative 
measurements reduce also errors from 
branching ratios and detection efficiency 
uncertainties.
\begin{table}[t]
\vspace{0.3cm}
\begin{center}
\begin{tabular}{|c|c|c|c|}
\hline
\multicolumn{4}{|c|}{Weak boson cross sections (LHC 14 TeV)} \\
\hline
\hline
 & \multicolumn{3}{|c|}{$\sigma \times BR$}\\
\hline
Reaction         & MRS(A)& CTEQ 2L & GRV 94 HO \\
\hline
$u \bar{d} \rightarrow W^{+} \rightarrow \ell^{+} \nu$
&20.18 nb  & 17.32 nb & 21.58 nb \\
\hline
$d \bar{u} \rightarrow W^{-} \rightarrow \ell^{-} \bar{\nu}$
&14.24 nb  & 12.63 nb & 15.40 nb \\
\hline
$ u\bar{u} (d\bar{d}) \rightarrow Z^{0} \rightarrow \ell^{+} \ell^{+}$
& 3.246 nb  & 2.854 nb & 3.456 nb \\
\hline
$q\bar{q}\rightarrow (Z^{*},\gamma^{*}) \rightarrow \ell^{+} \ell^{+}$ 
($M_{\ell\ell} = 150-200$ GeV)
& 9.71 pb  & 8.98 pb & 10.26 pb \\
\hline
$q \bar{q} \rightarrow W^{+}W^{-} \rightarrow \ell^{+} \nu 
\ell^{-} \bar{\nu}$
& 3.53 pb  & 3.30 pb & 3.63 pb \\
\hline
\end{tabular}\vspace{0.3cm}
\end{center}
\caption{Examples of estimated weak boson production cross sections
at the LHC for three different sets of structure functions
using PDFLIB and PYTHIA programs \cite{mrsa,pythia}.
In all cases the leptonic branching 
ratios into electrons and muons are included.
}
\end{table}

For the following studies, the MRS(A) structure 
function set \cite{mrsa}
is used as an example and reference system.   
Figure 1a shows the expected rapidity distribution 
of $W^{+}$ and $W^{-}$, which  
directly reflect
the difference between the $x$ distributions of the 
$u$, $d$ valence quarks and the sea quark or antiquarks.
For small $W^{\pm}$ rapidities, corresponding to $x_{1,2}$ values of 
$\approx 6\times 10^{-3}$, most $W^{\pm}$ originate from 
the annihilation of sea quark--antiquarks and only small 
differences between $W^{+}$ and $W^{-}$ are expected.
For larger rapidities the $W^{\pm}$ originate from 
the annihilation of quarks and antiquarks with 
very different $x$ values. For example, to produce 
a $W^{\pm}$ at a rapidity of about 2.5, one finds the 
corresponding $x_{1,2}$ values of the quark and antiquark to be 
$x_{1} \approx 0.1$ and $x_{2} \approx 3 \times 10^{-4}$.  
As the proton is made of 
two valence $u$ quarks and one valence $d$ quark
the $W^{+}$ production is much more 
likely than the $W^{-}$ production at large rapidities.

Figure 1b shows the pseudorapidity 
distributions of the charged leptons originating from 
the $W^{\pm}$ decays. Because of the 
V--A interaction, the differences between 
the pseudorapidity distributions of  
$\ell^{+}$ and $\ell^{-}$ especially at large $\eta$ values
are larger than the ones for the $W^{+}$ and $W^{-}$.
The reason is that 
the left handed lepton ($\ell^{-}$) is emitted preferentially
in the direction of the incoming quark and the right handed 
antilepton ($\ell^{+}$) is emitted opposite to the quark 
direction. Thus the observable charged lepton pseudorapidities
reflect not only the $x$ distributions of 
quarks and antiquarks but allow also to some extent 
a distinction 
between valence and sea quarks at a given $x$.

\begin{figure}[htb]
\begin{center}\mbox{
\epsfig{file=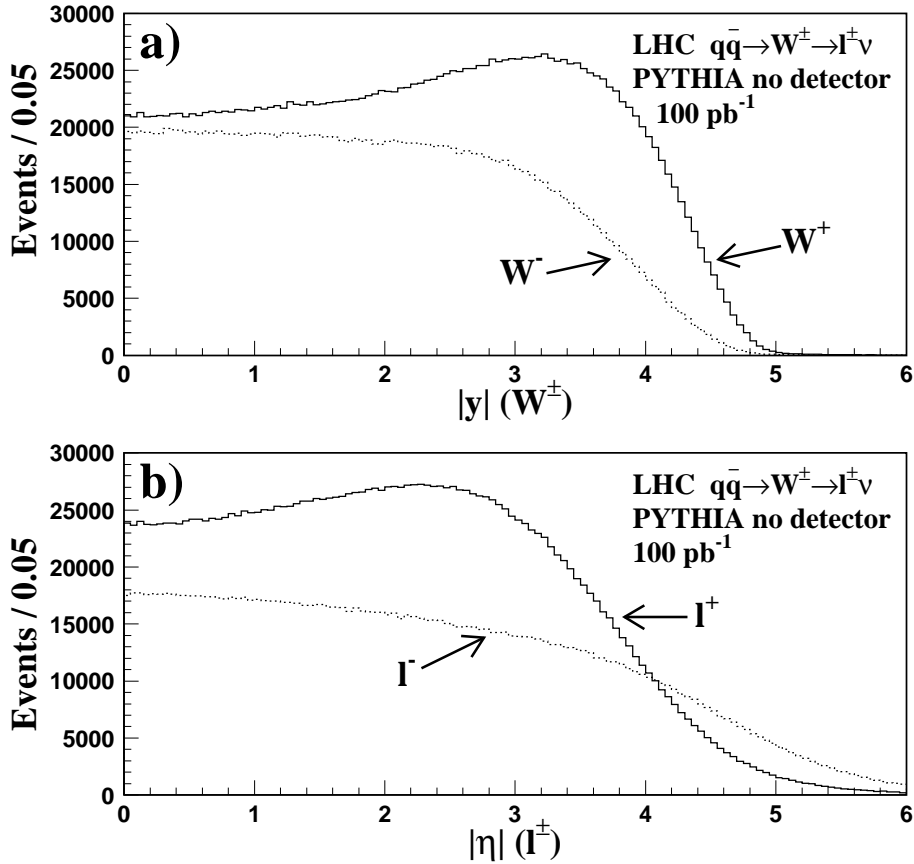,
height=13 cm,width=13cm}}
\end{center}

\caption[fig1]
{Expected rapidity (pseudorapidity) distribution of
$W^{\pm}$ (a) and $\ell^{\pm}$ (b) originating from the reaction  
$q\bar{q} \rightarrow W^{\pm} \rightarrow \ell^{\pm} \nu$ 
at the LHC ($\sqrt{s}=14$ TeV and the MRS(A) 
structure functions \cite{mrsa}). The assumed luminosity 
of 100~pb$^{-1}$ corresponds to about one day of data taking with 
a luminosity of $10^{33}~sec^{-1}~cm^{-2}$. 
}
\end{figure}

As discussed in the introduction, we want to demonstrate 
that the dynamics and event rates of weak boson production 
at the LHC accurately constrain the quark and antiquark 
structure functions and their corresponding luminosity. 
For this purpose simple event selection criteria are used.
These criteria closely follow the design characteristics of the 
proposed CMS experiment \cite{CMS}. In detail 
the following lepton selection criteria are used:
\begin{itemize}
\item
Electrons and Muons are required to have 
$p_{t} > 30$ GeV within a pseudorapidity of 
$|\eta| < 2.4$\footnote{Despite the 
interest in the very forward region,  
lepton detection up to much larger $|\eta|$ values
appears to be very difficult.}.
\item
In order to select only isolated leptons, the 
transverse energy deposit from other particles 
with $p_{t} > 0.5$ GeV and $|\eta| < 3$,
found within a cone R around the lepton  
($R \equiv \sqrt{\Delta \phi^{2} + \Delta \eta^{2}} < 0.6$),
should be smaller than 5 GeV.
\item
To reduce possible backgrounds from heavy quark decays 
and to reject high $p_{t}$ $W^{\pm}$ and $Z^{0}$ production
due to initial state radiation, events with reconstructed jets
with $E_{t} > 20$ GeV are removed.
The jet momentum vector is reconstructed in a cone $R < 0.6$
including all stable particles, 
with $p_{t} > 0.5$ GeV and $|\eta| < 3$.  
\end{itemize}

Using these charged lepton selection criteria, 
$pp \rightarrow W^{\pm} \rightarrow \ell^{\pm} \nu$ 
events are required to have 
exactly one isolated charged lepton with $30 < p_{t} < 50$ GeV. 
The resulting $p_{t}$ spectra of $\ell^{\pm}$ 
and their pseudorapidity distributions are shown
in figure 2a and 2b respectively.
The used kinematic and geometric event selection criteria, 
result in an event detection efficiency of 
about 25\% for $W^{+} \rightarrow \ell^{+} \nu$, 
and about 28\% for $W^{-}\rightarrow \ell^{-} \bar{\nu}$.
 
\begin{figure}[htb]
\begin{center}\mbox{
\epsfig{file=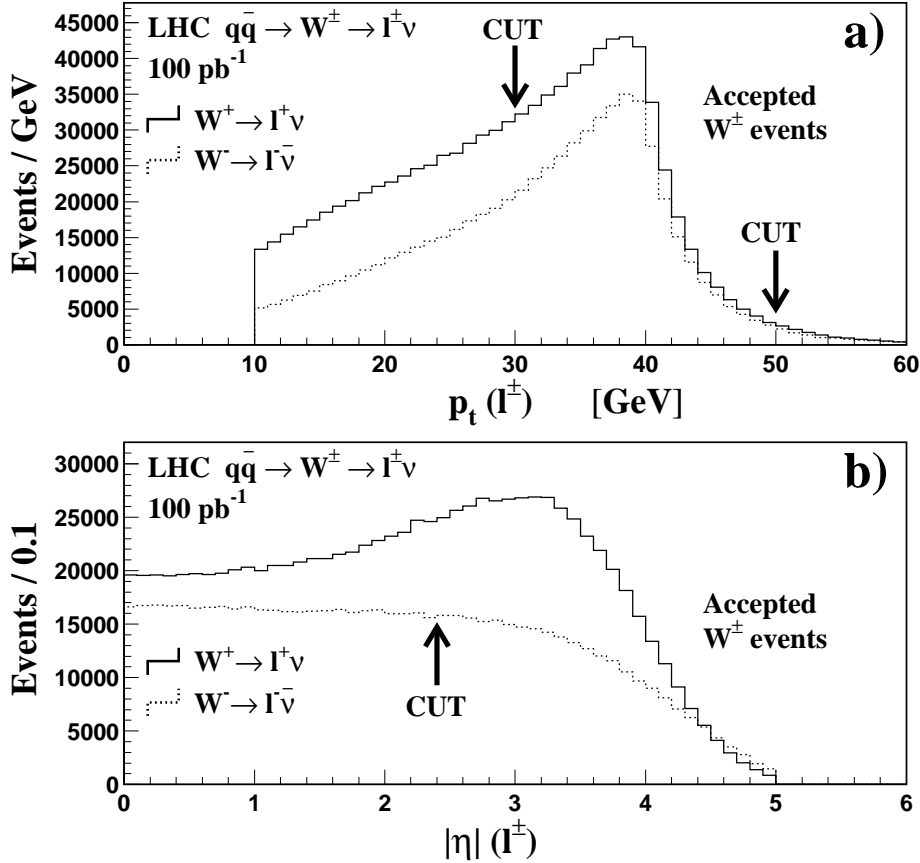,
height=13 cm,width=13cm}}
\end{center}
\caption[fig2]
{The observable (a)
charged lepton $p_{t}$ and (b) pseudorapidity $\eta$
distributions 
originating from the reaction  
$q\bar{q} \rightarrow W^{\pm} \rightarrow \ell^{\pm} \nu$ 
at the LHC ($\sqrt{s}=14$ TeV and the MRS(A) 
structure function \cite{mrsa}) 
including all selection criteria discussed in the text. 
}
\end{figure}

To select events of the type 
$pp \rightarrow Z^{0} \rightarrow \ell^{+}\ell^{-}$ 
the presence of a pair of isolated leptons with  
opposite charge ($e^{+}e^{-}$ or $\mu^{+}\mu^{-}$), with 
a $m_{\ell^{+}\ell^{-}} = m_{Z^{0}} \pm 2$ GeV.
In addition, the opening angle between the two leptons in the 
plane transverse to the beam has to 
be larger than $135^{\circ}$ and $p_{t}(Z^{0}) < 20$ GeV. 
As these dilepton events are usually considered 
to be background free, $Z^{0}$ events 
with large $p_{t}$ can be used to 
constrain the $x$ distribution of gluons,
as will be discussed later. 

Events of the type 
$q\bar{q} \rightarrow (Z^{*}, \gamma^{*}) \rightarrow \ell^{+}\ell^{-}$
with dilepton masses above 100 GeV have a much lower rate.
However, these events  
can be used to study the $Q^{2}$ evolution, up to 
masses where the neutral current 
sector is well understood (e.g. up to 
masses of about 200 GeV). At least up to these dilepton masses,
a measurement of the lepton forward backward charge asymmetry,
following the method of reference \cite{lhcafb}, constrains 
the ratio of valence and sea $u$ and $d$ quarks 
at different $x$ values.

Using the above kinematical and geometrical event selection criteria, 
the efficiency to detect both leptons from $Z^{0}$ decays
is about 16\%, and increases to about 23\% for 
dilepton masses in the range between 150--200 GeV.

\section {Sensitivity to the $q$ and $\bar{q}$ structure functions}

We now study the effects 
of different structure function parametrisations
on the measured $\ell^{\pm}$ pseudorapidity distributions 
($q\bar{q} \rightarrow W^{\pm} \rightarrow \ell^{\pm} \nu$),  
and on the reconstructed $Z^{0}$ rapidity distribution
($q\bar{q} \rightarrow Z^{0} \rightarrow \ell^{+} \ell^{-}$).

At the LHC,
in contrast to proton--antiproton colliders,
the antiquarks have to come from the sea.
Thus, the pseudorapidity distribution of the positive 
charged leptons, $u \bar{d} \rightarrow W^{+} \rightarrow \ell^{+} \nu$
contains the information about the sea $\bar{d}$ quarks
and the valence or sea $u$ quarks. 
The negative charged leptons,
$d \bar{u} \rightarrow W^{-} \rightarrow \ell^{-} \bar{\nu}$
carry consequently the information about 
the sea $\bar{u}$ quarks and the valence or sea $d$ quarks. 
The rapidity distribution of charged lepton pairs,  
from $Z^{0}, (Z^{*},\gamma^{*}) \rightarrow 
\ell^{+}\ell^{-}$, provide the information about  
the sum of sea $\bar{u}$ and $\bar{d}$ quarks
and the corresponding valence and sea quarks.
Consequently, the combination of the different 
observable lepton pseudorapidity 
distributions should provide some sensitivity to 
the $u$, $d$, $\bar{u}$ and $\bar{d}$ parton densities
over a large $x$ range.
   
This sensitivity is first investigated by comparing the 
weak boson production using two quite similar structure function sets, 
MRS(A) and MRS(H) \cite{mrsa}. 
The main difference between these 
two sets lies in the $x$ parametrisation for the light 
sea quarks. While the older MRS(H) set uses u, d flavor symmetric 
sea distributions, the MRS(A) set includes a fine tuning 
of the sea quark parton densities with 
some isospin symmetry breaking, required 
to describe the observation of Drell-Yan asymmetries  
of $A_{DY} = (\sigma_{pp} - \sigma_{pn}) / 
(\sigma_{pp} - \sigma_{pn})$ from the NA 51 experiment \cite{na51}.

Figure 3a shows the ratio of $\sigma(W^{+}) / \sigma(W^{-})$
as a function of the charged lepton pseudorapidity
for the two structure function sets. The different 
parametrisations thus lead, depending on the 
lepton pseudorapidity, to a cross section variation of up 
to about 10\%. The double ratio 
MRS(H)[$\sigma(W^{+}) / \sigma(W^{-})$]/
MRS(A)[$\sigma(W^{+}) / \sigma(W^{-})$] is shown 
in figure 3b. The differences of about 5-10\% 
between the two sets should be compared with the statistical 
precision, which is smaller than 1\% per bin for 
an integrated luminosity of only 100 pb$^{-1}$.
Furthermore, both sets of structure 
functions predict almost identical $Z^{0}$ cross sections.
The ratio between the $Z$ cross sections from the two 
sets, MRS(H)[$\sigma(Z^{0})$]/ MRS(A)[$\sigma(Z^{0})$], is 
also shown in figure 3b.  
Combining the obtainable information from $W^{+}$, $W^{-}$ 
and $Z^{0}$ production, the ``fine tuned" 
isospin splitting of $u$ and $d$ sea quarks between MRS(A)
and MRS(H) should be detectable with good accuracy.

\begin{figure}[htb]
\begin{center}\mbox{
\epsfig{file=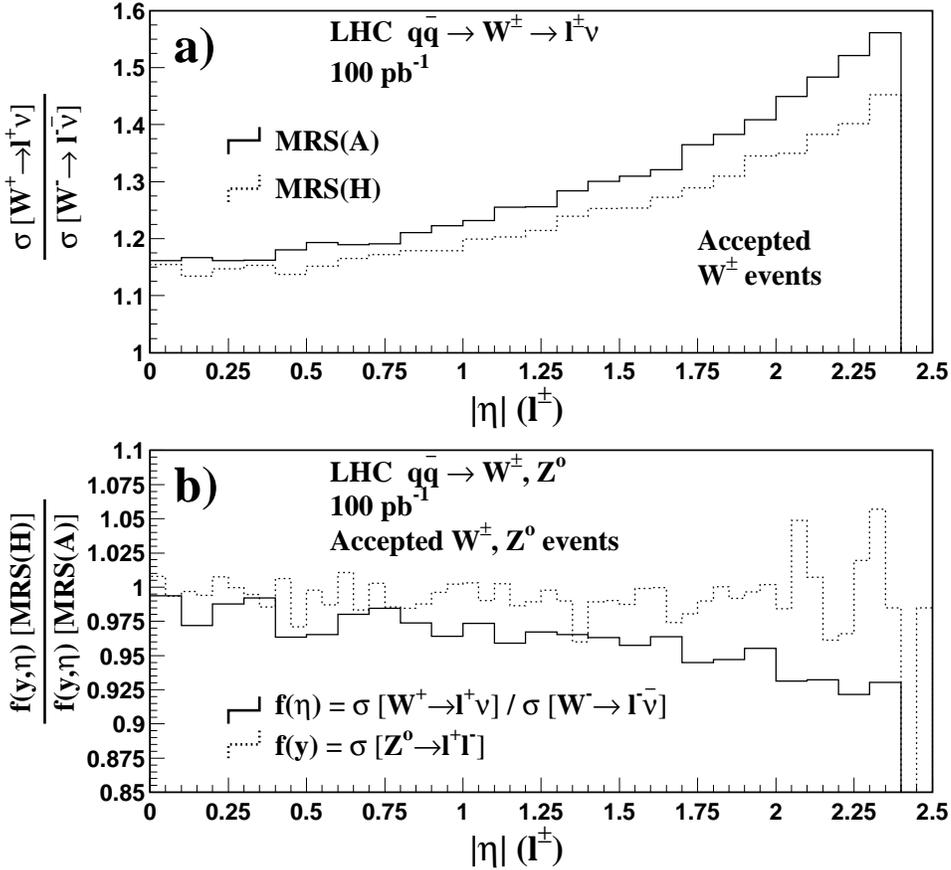,
height=13 cm,width=13cm}}
\end{center}
\caption[fig3]
{a) The detected charged lepton cross section ratio, 
$\sigma(\ell^{+} \nu)/\sigma(\ell^{-} \bar{\nu})$, originating 
from the reaction 
$q\bar{q} \rightarrow W^{\pm} \rightarrow \ell^{\pm} \nu$
as a function of the lepton pseudorapidity for  
the MRS(H) and MRS(A) structure function parametrisation. 
b) The relative changes of the 
charged lepton ratios of 3a) between  
the MRS(H) and MRS(A) parametrisations 
and also the cross section ratio of the $Z^{0}$ production 
using both parametrisations.
}
\end{figure}
 
We have shown that 
the weak boson rapidity distributions are sensitive to 
small differences between the  
$x$ distribution of $u$ and $d$ sea quarks and antiquarks.
We now go one step further and study how well $q$ and $\bar{q}$
structure function can be constrained from 
the observable weak boson rapidities.
For this purpose the different $\ell^{\pm}$ cross sections are 
studied relative to a reference structure function,
arbitrarily chosen to be the MRS(A) set.

The fraction of weak bosons which are produced
from the annihilation of valence quarks and low $x$ antiquarks 
increases strongly with increasing rapidity. 
The valence quark $x$ distribution is already
quite well constrained. The main difference between the 
various structure functions comes from 
the sea $q$ and $\bar{q}$ parametrisations especially at low $x$. 
Thus precise measurements of  
the charged lepton pseudorapidity distributions
from $W^{\pm}$ decays 
and the rapidity distribution of the $Z^{0}$ events 
constrain the low $x$ domain of $\bar{u}$ and $\bar{d}$.

The sensitivity of the measurable lepton rapidity distribution
to the different sets of structure functions is shown in figure
4 and figure 5. Figure 4 shows the observable ratio of the 
$\ell^{+}$ to $\ell^{-}$ event rates for three different 
sets of structure functions and an integrated luminosity of
100 $pb^{-1}$. The difference between 
the various low $x$ sea quark parametrisations are thus reflected
in the observable lepton pseudorapidities.
Consequently, the shape of the $\ell^{\pm}$ pseudorapidity 
distributions provide a strong constraint on the 
underlying $x$ distribution of quarks and antiquarks
with $x$ between $\approx 3 \times 10^{-4}$ and  
$\approx 10^{-1}$.

\begin{figure}[htb]
\begin{center}\mbox{
\epsfig{file=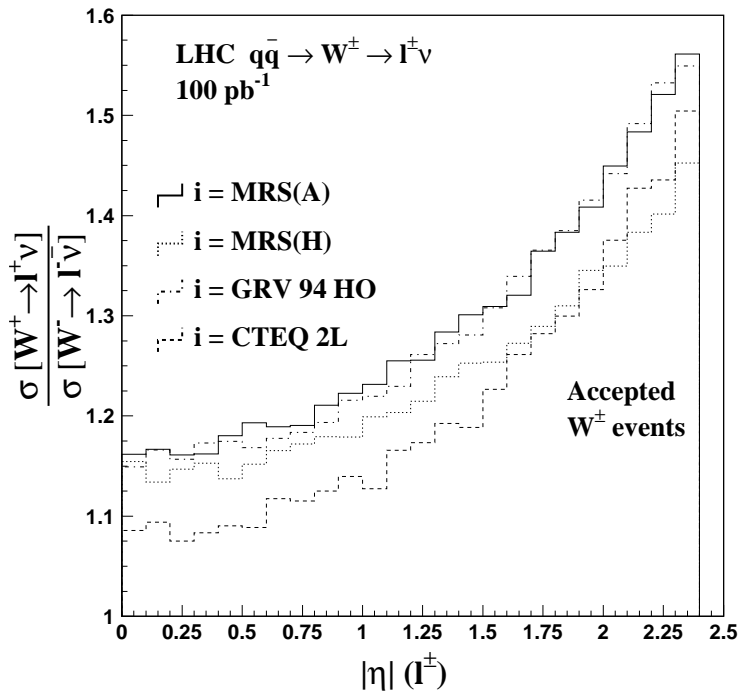,
height=10 cm,width=10cm}}
\end{center}
\caption[fig4]
{The ratio of the accepted 
cross sections 
$\sigma(u\bar{d}\rightarrow W^{+} \rightarrow \ell^{+} \nu)$
and $\sigma(d\bar{d} \rightarrow W^{-} \rightarrow \ell^{-} \bar{\nu})$
as a function of the lepton pseudorapidity
for four different structure functions \cite{mrsa}.}
\end{figure}
Figures 5a-c show the ratio of the 
predicted $\ell^{\pm}$ cross sections 
from different structure functions relative to the 
reference MRS(A) set. The statistical fluctuations 
shown in figures 5a and 5b correspond to the errors from  
roughly one day of data taking at the 
initial LHC luminosity of $10^{33} sec^{-1} cm^{-2}$. 
The expected $Z^{0}$ event rates are roughly a factor of 10 smaller 
and the errors shown in figure 5c correspond to about 10 days of 
data taking. 
\begin{figure}[ht]
\begin{center}
%\mbox{
\epsfig{file=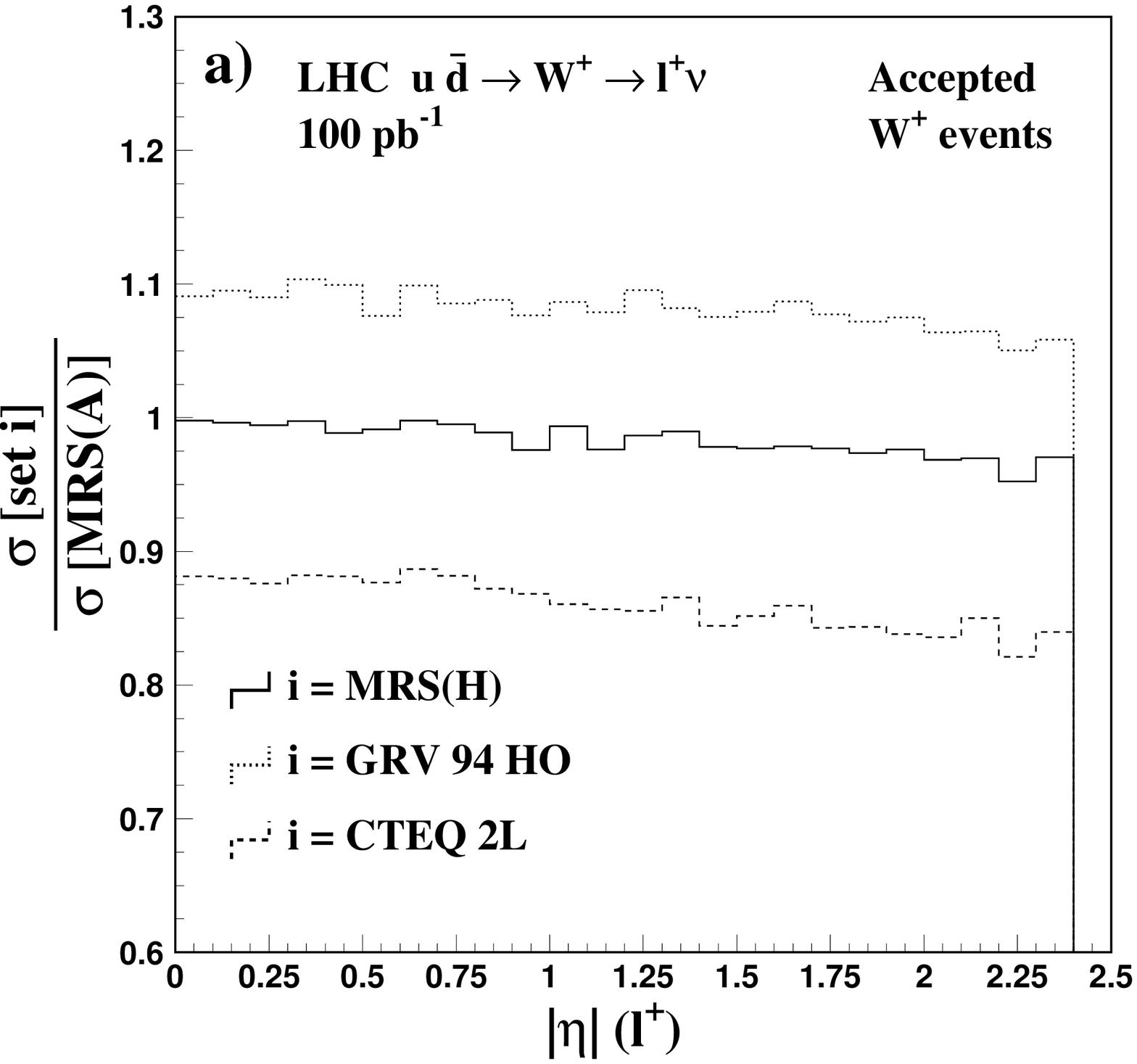,
height=8 cm,width=8cm}
\epsfig{file=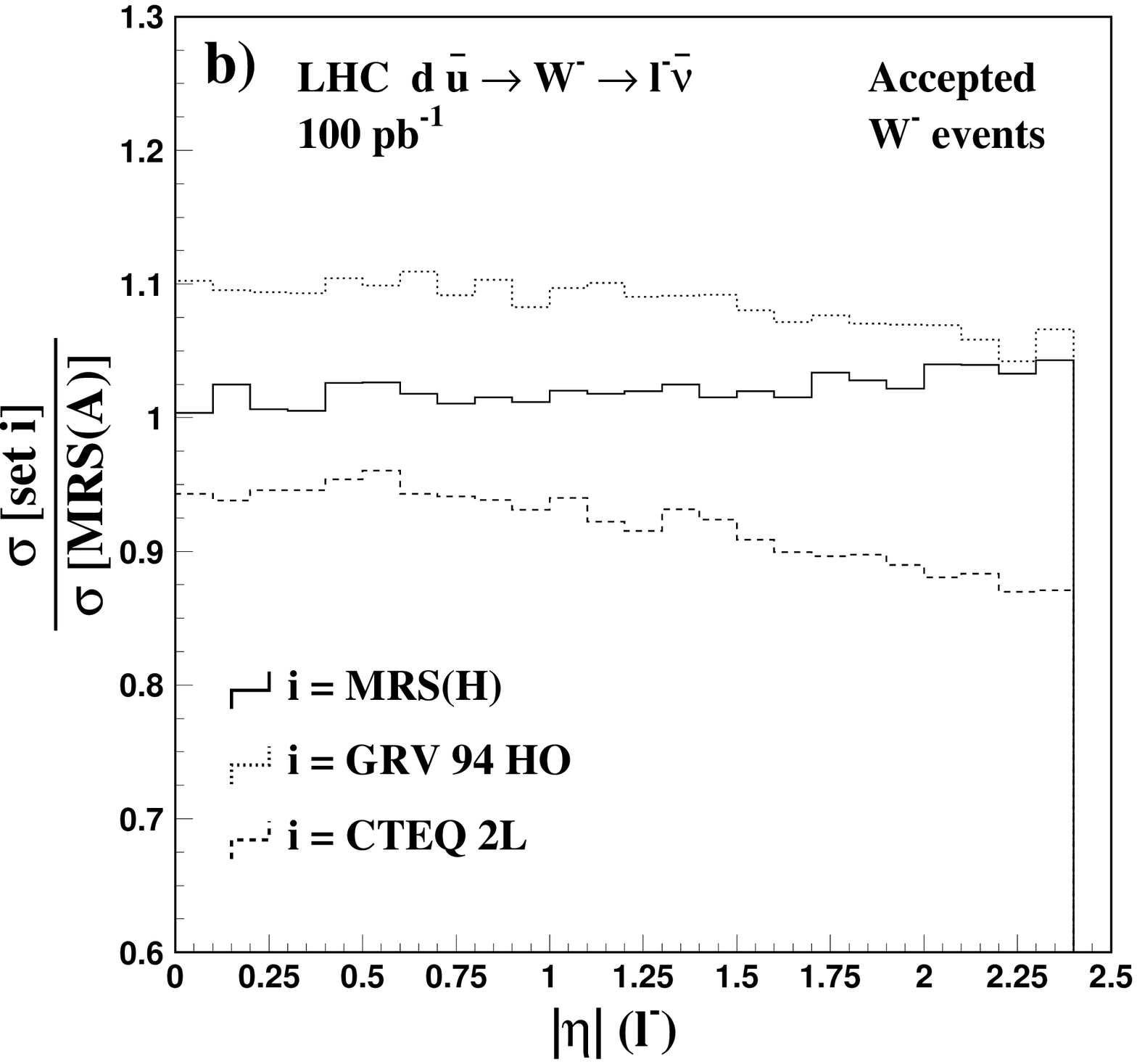,
height=8 cm,width=8cm}
%}
%\mbox{
\epsfig{file=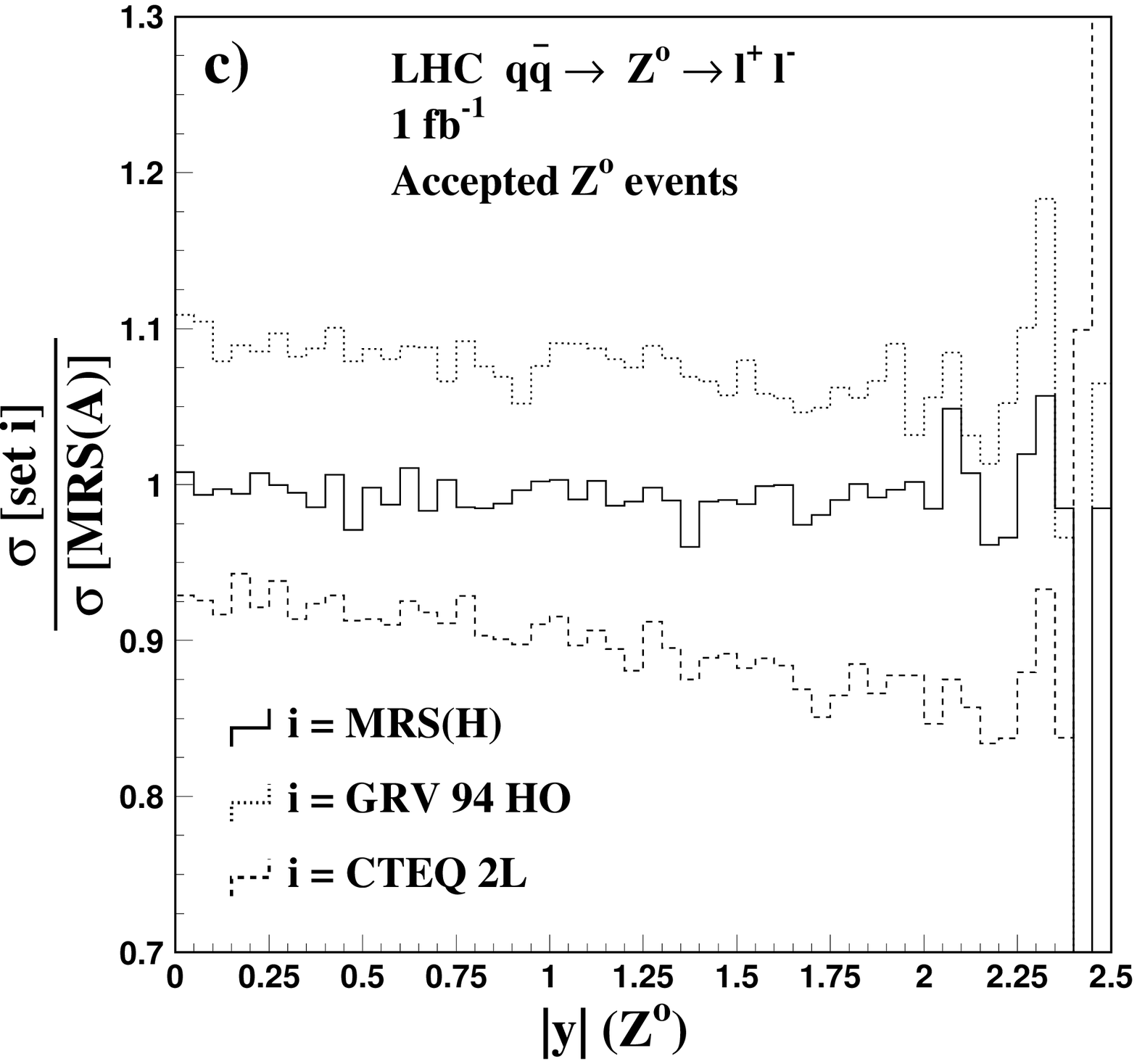,
height=8 cm,width=8cm}
%}
\end{center}
\caption[fig5]
{Rapidity dependence of the $\ell{^\pm}$ 
cross section predictions from different 
sets of structure functions relative to the one 
obtained from the MRS(A) parametrisation; a) for $\ell^{+}$,
b) for $\ell^{-}$ and c) for the reconstructed $Z^{0}$.
}
\end{figure}
\clearpage
Having demonstrated that the $\ell^{\pm}$ pseudorapidity 
distributions, originating from weak boson decays, 
are very sensitive to details of the quark and antiquark
$x$ distributions one can now relate the rate 
of $\ell^{\pm}$ events in a selected pseudorapidity 
interval to the quark and antiquark luminosity at the 
given $x$. Obviously, once the shape of the 
pseudorapidity distribution is accurately known,
the $\ell^{\pm}$ event rates 
need to be measured only for a small pseudorapidity interval.
For example, counting of $\ell^{\pm}$ events from the 
process $pp \rightarrow W^{\pm} \rightarrow \ell^{\pm} \nu $
could be restricted to 
the pseudorapidity range of $|\eta| < 0.5$.
Including all selection criteria one would 
observe roughly 150 000 ``clean" luminosity events, corresponding
to a statistical error of 0.3\%,
per day at the initial LHC luminosity ($\approx 100~pb^{-1}/day$).

Once the quark and antiquark luminosity 
at $Q^{2} \approx 10^{4} $ GeV$^{2}$ and in the $x$ range between
$\approx 5\times 10^{-4}$ and $10^{-1}$ are determined, 
accurate cross section predictions 
of other $q\bar{q}$ related processes should be possible.
This is studied for the reaction $ q\bar{q} \rightarrow W^{+}W^{-}$. 
The correlation between cross section predictions for 
single and pair production of weak bosons 
has been pointed out already in section 2 (see table 1).
For example, the total cross section predictions
for the process $pp \rightarrow W^{\pm}$ between 
the CTEQ 2L and the MRS(A) parametrisations 
differ by about 15\%. 
However, as we suggest to use the 
process $\sigma(q\bar{q} \rightarrow W^{\pm})$ 
as a reference process, one has to 
relate the cross section of for example  
$\sigma(q\bar{q} \rightarrow W^{+}W^{-})$ to the 
reference 
reaction $\sigma(q\bar{q} \rightarrow W^{\pm})$.
Comparing now the prediction for the
relative cross sections between 
CTEQ 2L and the MRS(A) one finds that the difference 
is reduced to $\approx$ 7.5\%.

As a next step, the parametrisations 
of the $q, \bar{q}$ structure functions, 
especially at low $x$, should 
be adjusted such that the observed $\ell^{\pm}$ 
pseudorapidity distributions are described. 
As the final experimental accuracy for the lepton 
pseudorapidity distributions will be limited by systematics, 
the limitations of the structure function ``fine tuning" 
are difficult to estimate. It is nevertheless worth pointing 
out that neither the $\ell^{\pm}$ momentum and 
charge determination nor 
differences between $\ell^{+}$ and $\ell^{-}$ detection
are expected to be problematic. Furthermore, backgrounds
from different sources and efficiency uncertainties
can be controlled by the simultaneous  
analysis of the $W^{\pm}$ and $Z^{0}$ events with isolated 
electrons and muons. We therefore do not 
expect any principle 
problem of measuring the shape and the rate 
of the charged lepton pseudorapidity 
distribution with a $\pm$1\% accuracy. 
Thus even small differences for the sea quark
parametrisation,
like those between MRS(A) and MRS(H), 
as shown in figure 3a and b,
should be detectable.
One could thus use the difference in cross section
for the two sets as a 
pessimistic limitation of the proposed method.
Differences between relative cross section 
predictions for different $q\bar{q}$ scattering processes
and the two parametrisations indicate therefore 
the size of the remaining uncertainties.
For example the cross section ratios 
$\sigma(q\bar{q} \rightarrow W^{+}W^{-})/
\sigma(q\bar{q} \rightarrow W^{\pm})$ are 
$4.74\times 10^{-4}$ for MRS(A) and 
$4.76\times 10^{-4}$ for MRS(H).
Other $q\bar{q}$ scattering processes
like $\sigma(q\bar{q} \rightarrow W^{\pm}Z^{0})/
\sigma(q\bar{q} \rightarrow W^{\pm})$ 
show similar stability with 
predicted ratios of 
$1.78\times 10^{-4}$ for MRS(A) and 
$1.79\times 10^{-4}$ for MRS(H).

Following the above procedure, i.e.
constraining the $q,\bar{q}$ structure functions
and the corresponding parton luminosities, the 
event rate of weak boson pair production appears to be 
predictable with an accuracy of  
at least $\pm$1\%.

\section{Outlook and conclusions} 

A new approach to the LHC luminosity measurement 
demonstrates that the 
$x$ distributions of valence and sea quarks and their 
corresponding parton luminosities can be
constrained very accurately, using the 
$\ell^{\pm}$ pseudorapidity distributions 
from the decay of weak bosons.
It is also shown that this method 
leads to very accurate rate predictions of other 
$q\bar{q}$ scattering processes.
For example, the strong correlation between 
the weak boson pair production and the single boson 
production leads to an estimated experimental luminosity 
accuracy at the $\pm$1\% level. This should be compared 
to the often considered optimistic goal of $\pm$5\% accuracy.

We have not investigated the achievable theoretical 
accuracies, but believe that many theoretical uncertainties, 
like the $\alpha_{s}(Q^{2})$ 
uncertainties or still unknown higher order QCD corrections,
contribute in very similar ways to the single and pair production 
of weak bosons. Furthermore, the experimental possibility
to measure the $x$ distributions of
sea and valence quarks
and the corresponding luminosities to within $\pm$1\%
should encourage our theoretical colleagues to match 
this experimental accuracy.

Finally, we argue that the gluon $x$ distribution and the 
corresponding gluon luminosity can also be constrained
in a similar way from accurate 
measurements of the rapidity distribution
of gluon dominated scattering processes. 
In fact, as the $q, \bar{q}$ luminosity can accurately 
be measured from the weak boson rapidity
distribution, the rapidity distribution of gluon dominated 
scattering processes has only to be measured 
relative to the weak boson rapidity distributions.
Once the gluon $x$ distribution is known relative
to the $x$ distribution of quarks, 
the weak boson rate also provides
the luminosity monitor for gluon related signal and 
background processes. 

The possible experimental accuracy thus depends
mainly on how accurate the rapidity and $Q^{2}$ distributions 
of gluon related scattering processes can be measured. 
A very clean signature with a well measurable $Q^{2}$
is much more important than a huge cross section.

Gluon related scattering processes are 
$g g \rightarrow X$ and $g q(\bar{q}) \rightarrow X$.
As these processes involve jets, 
measurement problems should be minimized by using 
processes with small backgrounds and 
well measurable $p_{t}$.
Candidates for such processes are
high $p_{t}$ events with one or more jets and an isolated 
$\gamma$ or a $Z^{0} (\rightarrow \ell^{+} \ell^{-}$). 
As the energy and momentum of isolated 
photons and leptons can be measured very accurately,
the $p_{t}$ of the jets, assuming transverse 
momentum conservation, can also be determined. Thus, 
the observables are well measured and should provide
accurate $Q^{2}$ measurements.

The production of events with isolated high $p_{t}$
photons or $Z^{0}$ are dominated by 
$g q \rightarrow \gamma (Z^{0}) q$,
and $q \bar{q} \rightarrow \gamma (Z^{0}) g$.
As shown in table 2, the expected cross sections 
for these reactions, including the branching ratios
$Z^{0} \rightarrow \ell^{+}\ell^{-}$, and  
relatively high $p_{t}$ of $\gamma$ and $Z^{0}$ 
are still quite large.
Furthermore, the calculable background corrections from the 
process $q \bar{q}$ are expected to be small 
as the cross sections 
are dominated by the $q g$ scattering process.
 
\begin{table}[t]
\vspace{0.3cm}
\begin{center}
\begin{tabular}{|c|c|c|c|}
\hline
\multicolumn{4}{|c|}{LHC 14 TeV for MRS(A) and PYTHIA 5.7} \\
\hline
reaction & $\sigma$ [pb]  & $\sigma$ [pb] & $\sigma$ [pb] \\
  & $50 GeV <p_{t} < 100$ GeV& $100 GeV <p_{t} < 200$ GeV & 
$p_{t} > 200$ GeV \\
\hline
$q \bar{q} \rightarrow Z^{0}(\rightarrow \ell^{+}\ell^{-}) g$
&  36.4  & 6.01 & 0.71 \\
\hline
$q g \rightarrow Z^{0}(\rightarrow \ell^{+}\ell^{-}) q$
& 150  & 34.8 & 4.08 \\
\hline
\hline
$q q \rightarrow \gamma g$
& 717  & 74.5 & 7.45 \\
\hline
$q g \rightarrow \gamma q$
&6590  & 615 & 49.3 \\
\hline
\end{tabular}\vspace{0.3cm}
\end{center}
\caption{PYTHIA cross section estimates for high $p_{t}$
final states of the type 
$Z^{0} (\rightarrow \ell^{+}\ell^{-}) q (g)$ 
and $\gamma q(g)$.
}
\end{table}

Previous studies of $\gamma$--jet final 
states have shown that jet events with 
isolated $\pi^{0}$'s provide 
a considerable background \cite{gammajet}.
This large background will therefore limit 
the achievable accuracy of such a final state. 
However, the leptonic $Z^{0}$ decays 
provide an excellent signature and should allow 
the selection of essentially background free 
$Z^{0}$--jet events.
We are not aware of any detailed LHC study which demonstrates
that this process can indeed be measured with accuracies of a
few \%,
but see no obvious reason why the rapidity distribution 
of the clean $Z^{0}$--jet events can not be measured 
with an accuracy close to $\pm$1\%.

Unfortunately, the accurate $Q^{2}$ determination due to  
inherent uncertainties of jet energy measurements 
especially at large rapidities will probably limit 
the interpretation of the observable rapidity distribution 
with respect to the gluon $x$ distribution.
Nevertheless, such direct measurements of the gluon 
structure function will provide the highest possible 
accuracy for the $x$ distribution of gluons and 
might eventually lead to cross section predictions 
with \% accuracies for other gluon related scattering processes.

To summarise, we have shown that 
the rapidity distributions of $W^{\pm}$ and $Z^{0}$ events
at the LHC provide directly and accurately the 
$x$ distributions of quarks and antiquarks. Their rates 
are thus a measure of the corresponding parton luminosities. 
We have shown that such an approach might eventually lead 
to perhaps 1\% accurate cross section predictions of $q \bar{q}$ 
related scattering processes, 
like $q\bar{q} \rightarrow W^{+}W^{-}$ at the LHC.
We suggest that the detailed measurement of the rapidity 
distribution of the process $q g \rightarrow Z q$ might 
provide similar accuracies for the gluon structure function
and the corresponding gluon luminosity.

\vspace{2.cm}
% \acknowledgments
{\bf \large Acknowledgements}

We would like to thank H. Dreiner for many detailed 
discussions and comments on the manuscript. We are also 
grateful to M. Spira for several detailed discussions and 
to R. G. Roberts for his comments.

\end{document}